\documentclass[conference]{IEEEtran}
\IEEEoverridecommandlockouts
\usepackage{cite}
\usepackage{tabularx}
\usepackage{diagbox}
\usepackage{amsmath,amssymb,amsfonts}
\usepackage{algorithmic}
\usepackage{graphicx}
\usepackage{booktabs}
\usepackage{textcomp}
\usepackage{hyperref}
\usepackage{xcolor}

\def\BibTeX{{\rm B\kern-.05em{\sc i\kern-.025em b}\kern-.08em
    T\kern-.1667em\lower.7ex\hbox{E}\kern-.125emX}}

\makeatletter
\newcommand{\linebreakand}{%
  \end{@IEEEauthorhalign}
  \hfill\mbox{}\par
  \mbox{}\hfill\begin{@IEEEauthorhalign}
}

\begin{document}
\title{Knowledge Distilled Ensemble Model for sEMG-based Silent Speech Interface}



\author{\IEEEauthorblockN{Wenqiang Lai\textsuperscript{\dag}\thanks{\dag: All authors have the same contribution. The co-first authorship order follows our group name \textit{LYMSY}. This work has been submitted to the IEEE Region 8 Student Paper Contest 2023 for possible publication. Copyright may be transferred without notice, after which this version may no longer be accessible.}}
\IEEEauthorblockA{\textit{Electrical \& Electronic Eng} \\
\textit{Imperial College London}\\
London, UK \\
wenqiang.lai21@imperial.ac.uk}
\and
\IEEEauthorblockN{Qihan Yang\textsuperscript{\dag}}
\IEEEauthorblockA{\textit{Electrical \& Electronic Eng} \\
\textit{Imperial College London}\\
London, UK \\
qihan.yang21@imperial.ac.uk}
\and
\IEEEauthorblockN{Ye Mao\textsuperscript{\dag}}
\IEEEauthorblockA{\textit{Electrical \& Electronic Eng} \\
\textit{Imperial College London}\\
London, UK \\
ye.mao21@imperial.ac.uk}

\linebreakand
\IEEEauthorblockN{Endong Sun\textsuperscript{\dag}}
\IEEEauthorblockA{\textit{Electrical \& Electronic Eng} \\
\textit{Imperial College London}\\
London, UK \\
endong.sun21@imperial.ac.uk}
\and
\IEEEauthorblockN{Jiangnan Ye\textsuperscript{\dag}}
\IEEEauthorblockA{\textit{Electrical \& Electronic Eng} \\
\textit{Imperial College London}\\
London, UK \\
jiangnan.ye21@imperial.ac.uk}
}
\maketitle
\begin{abstract}
Voice disorders affect millions of people worldwide. Surface electromyography-based Silent Speech Interfaces (sEMG-based SSIs) have been explored as a potential solution for decades. However, previous works were limited by small vocabularies and manually extracted features from raw data. To address these limitations, we propose a lightweight deep learning knowledge-distilled ensemble model for sEMG-based SSI (KDE-SSI). Our model can classify a 26 NATO phonetic alphabets dataset with 3900 data samples, enabling the unambiguous generation of any English word through spelling. Extensive experiments validate the effectiveness of KDE-SSI, achieving a test accuracy of 85.9\%. Our findings also shed light on an end-to-end system for portable, practical equipment.
\end{abstract}


\section{Introduction}
\label{sec:intro}
Normal communication is not always possible. According to a report from the American Speech-Language-Hearing Association (ASHA), nearly 40 million US citizens have communication disorders, which cost the US approximately $154-186$ a billion dollars annually. Diseases that lead to language impairments include brain injuries (e.g., aphasia, apraxia, and dysarthria) and voice disorders, where there are disturbances in the vocal folds or any other organ involved in voice production. Surface electromyographic signal-based SSIs (sEMG-based SSIs) are one of the standard solutions to voice disorder \cite{gonzalez2020silent}, which recognises speech from EMG signals recorded from speech-related facial muscles in a non-invasive manner (via the electrodes attached to the skin). The electrodes are easy to apply, requiring no medical supervision and certification. Most useful information in sEMG signals is in the frequency band between $15$Hz and $450$Hz~\cite{viitasalo1977signal}. 

The initial research of sEMG-based SSI dates back to the mid-1980s~\cite{Morse1989,sugie1985speech}. Until the early-2000s, the literature mainly focused on few words classification and achieved high accuracy~\cite{chan2001myo, Jorgensen2003}. However, limited vocabulary made those systems less usable in practice. Approaches to address this issue included dividing the word into sub-word units~\cite{walliczek2006sub} or learning the phonetic feature in a data-driven manner~\cite{schultz2010modeling}. Those methods promoted the sEMG-based SSI towards continuous speech recognition, but the methods were mainly based on traditional machine learning and manually engineered features (Decision Tree, Linear Discriminant Analysis, Gaussian Mixture Model), and words regenerated from dissembled syllabuses could be erroneous. Most recently, deep learning-based methods have thrived and significantly improved over conventional models~\cite{Wand2014}. AlterEgo, utilising CNN, proposed a product that did not require users explicitly mouth their speech with pronounced, apparent facial movements~\cite{kapur2018alterego}. Nevertheless, the currently available product usually relies on expensive sensors and are less affordable. 

In this work, we innovatively applied a new proposed deep-learning method to classify the International Radiotelephony Spelling Alphabet with a commercially off-the-shelf (COTS) device. The International Radiotelephony Spelling Alphabet, commonly known as the NATO (North Atlantic Treaty Organization) phonetic alphabet or ICAO (International Civil Aviation Organization) phonetic alphabet, is the most widely used set of clear code words for communicating the letters of the Roman alphabet over the phone or military radio \cite{alphabet}. Each ``code word" in the alphabet stands for its initial letter. The $26$ code words and their corresponding alphabetical symbol are listed in Table~\ref{Tab:alphabet}. With the help of the NATO alphabet, any word constructed from the $26$ Roman character can be spelt unambiguously.
\begin{table}[ht!]
\caption{NATO phonetic alphabet.}
\centering
\resizebox{!}{!}{
\begin{tabular}{@{}llllll@{}}
\toprule
 &  & \multicolumn{3}{l}{\textbf{Phonetic Alphabet}} &  \\ \midrule
A & - Alpha & J & - Juliet & S & - Sierra \\
B & - Bravo & K & - Kilo & T & - Tango \\
C & - Charlie & L & - Lima & U & - Uniform \\
D & - Delta & M & - Mike & V & - Victor \\
E & - Echo & N & - November & W & - Whisket \\
F & - Faxtrot & O & - Oscar & X & - X-Ray \\
G & - Golf & P & - Papa & Y & - Yankee \\
H & - Hotel & Q & - Quebec & Z & - Zulu \\
I & - India & R & - Romeo &  &  \\ \bottomrule
\end{tabular}
}
\label{Tab:alphabet}
\end{table}

\begin{figure*}[ht!]
\centerline{\includegraphics[width=\textwidth]{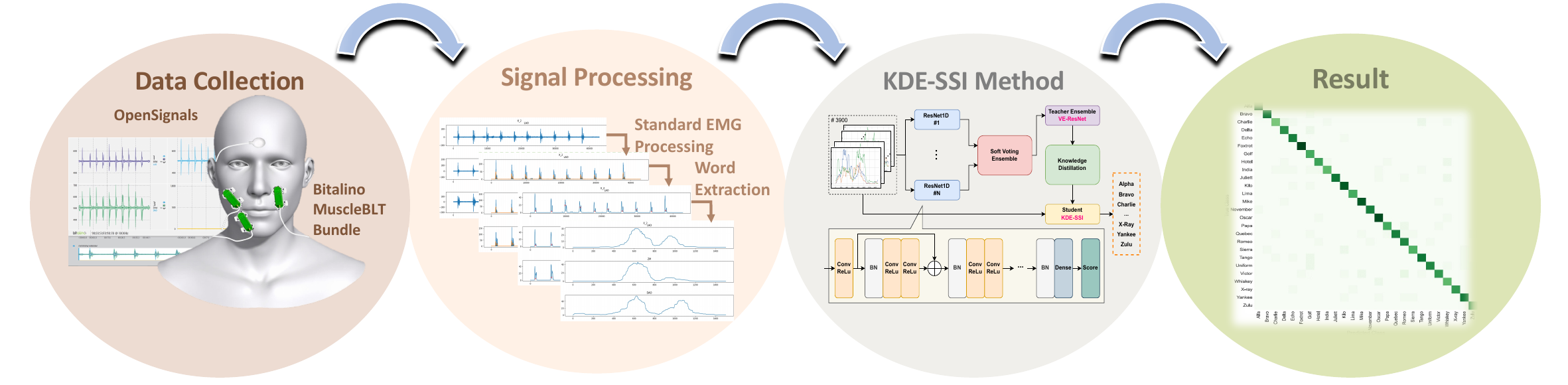}}
\caption{\textbf{The overall workflow of the project.} Data collection: collect $3900$ data samples of $3$ facial muscles from $5$ subjects using the Bitalino MuscleBLT bundle and OpenSignals software; Signal processing: standard EMG signal processing steps followed by word extraction; Model construction: using ResNet1D as the backbone, the ensemble model applies soft voting and achieves a better result than any single model; Result analysis: the performance of baseline model, ensemble model with a different number of backbones, and KDE-SSI with different temperatures are presented.}
\label{Fig:workflow}
\end{figure*}

The main contribution of our work is threefold: 1) construct a $26$ words NATO phonetic alphabet dataset ($3900$ data samples in total) from the facial sEMG signals of $5$ male subjects; 2) achieve $81.2\%$ test accuracy for the single model on the created dataset; 3) implement a Knowledge Distilled Ensemble Model for Silent Speech Interface (KDE-SSI), which is a lightweight convolutional network that efficiently extracts knowledge from a pre-trained voting ensemble ResNet model (VE-ResNet) while maintaining performance. It outperforms any backbones, achieving $85.9\%$ accuracy.

In the following sections, the article explains the technical details. The overall workflow is shown in Figure~\ref{Fig:workflow}. Section~\ref{sec:Dataset} introduces the dataset and collection procedure; section~\ref{sec:Processing} illustrates the signal processing pipeline; section~\ref{sec:Method} explains the method used, and section~\ref{sec:result_discussion} discusses the results obtained and their implications.

\section{Dataset}
\label{sec:Dataset}

We collected $30 \times 26=780$ samples for each of the five subjects, totalling $3900$ ($150$ per class). The hardware, sensor placement, and software used in this project are introduced below.

\subsection{Hardware and Sensor Placement}
The hardware used in data collection was the BITalino MuscleBIT bundle, containing a pre-assembled BITalino Core (onboard microprocessor, Bluetooth device and battery), four assembled sEMG sensors with prefixed electrode distance ($1.5$cm), one reference cable, a Bluetooth dongle, and pre-gelled self-adhesive disposable Ag/AgCl electrodes.

Three speech-related facial muscles were selected: \textit{levator anguli oris} (LAO), \textit{depressor anguli oris} (DAO), and \textit{zygomaticus major} (ZM). The electrodes were placed between the motor unit and the tendinous insertion of the muscle, with their longitudinal axes aligned with the midline of the muscle~\cite{de2002surface}.

\subsection{Software}
The software used in data collection was OpenSignals, an easy-to-use and versatile software suite for real-time bio-signals visualisation. It was compatible with the BITalino MuscleBIT bundle and communicates through Bluetooth.

Data was collected in $1000$ Hz (to satisfy the Nyquist sampling rate) from three channels simultaneously and the raw data was stored in the local computer in \texttt{H5} format.

\subsection{Data Collection Protocol}
Data were collected from $5$ male subjects aged from $22$ to $24$. Before the data collection, subjects sat comfortably in front of a desk with sensors attached to their faces. Each subject was asked to collect the data for the $26$ words in sequence. For each word, $3$ trials were carried out. In each trial, the word was repeated for $10$ times. The subjects should finish mouthing the word in $2$ seconds and rest another $2$ seconds before taking the next try. Hence, each trial lasted around $1$ minute, and the collection section for each subject lasted for around $2$ hours.

\section{Signal Processing}
\label{sec:Processing}
Before the data could be used to train the model, certain pre-processing steps were required. Two stages were involved: 1) the standard EMG signal processing, which eliminated the noises in the raw signal and acquired the enveloped-data; 2) word extraction, which extracted the $10$ words from the data stream.
\subsection{Standard EMG Signal Processing}
First, the raw EMG data was changed from the \texttt{H5} data to the \texttt{CSV} format. Then the following steps were taken:

\textbf{Zero-mean:} the mean was subtracted from each channel.

\textbf{Denoising:} wavelet denoising has proved effective in bio-signal processing since it has good frequency resolution at high frequencies; thus, the noise components in a signal can be isolated while important high-frequency transients can also be preserved~\cite{Hussain2009wavelet}. This project used Daubechies wavelets (db$2$) at decomposition level $4$ and soft minimax thresholding.

\textbf{Filtering:} as stated in the Section~\ref{sec:intro}, most useful sEMG signals are located between $15$-$28$ Hz to $400$-$450$ Hz. Most low-frequency noises come from the skin-electrode interface: the muscle movements underneath the skin cause the \textit{movement artefact noise}. Meanwhile, most of the high-frequency noises are random noises. Hence, a $10$th order Butterworth filter was implemented. The low-pass and high-pass filter corner frequencies were set to $400$ Hz and $20$ Hz, respectively.

\textbf{Rectification:} Full-wave rectification took the absolute value of all the negative values and turned them positive.

\textbf{Envelope:} RMS envelope was extracted from the raw data. It was calculated by computing the root-mean-square (RMS) value of the signal within a window that slides across the signal. The formula is shown in \eqref{Eq:RMS_envelope}. Here $x(n)$ is the rectified signal, $N_w$ is the window length which is $100ms\times1000Hz=100$, $L$ is the length of the data, and $e(n)$ is the envelope value calculated from the window located at time step $n$.
\begin{align}
\label{Eq:RMS_envelope}
    e(n) &= \sqrt{\frac{1}{N_w}\sum_{N \in -{N_w}/2}^{{N_w}/2} x(n+N)^2}, \\
    n \in[{N_w}/2, &{N_w}/2+1, ... L-{N_w}/2-1, L-{N_w}/2] \notag
\end{align}

\begin{figure}[ht!]
\centerline{\includegraphics[width=0.9\columnwidth]{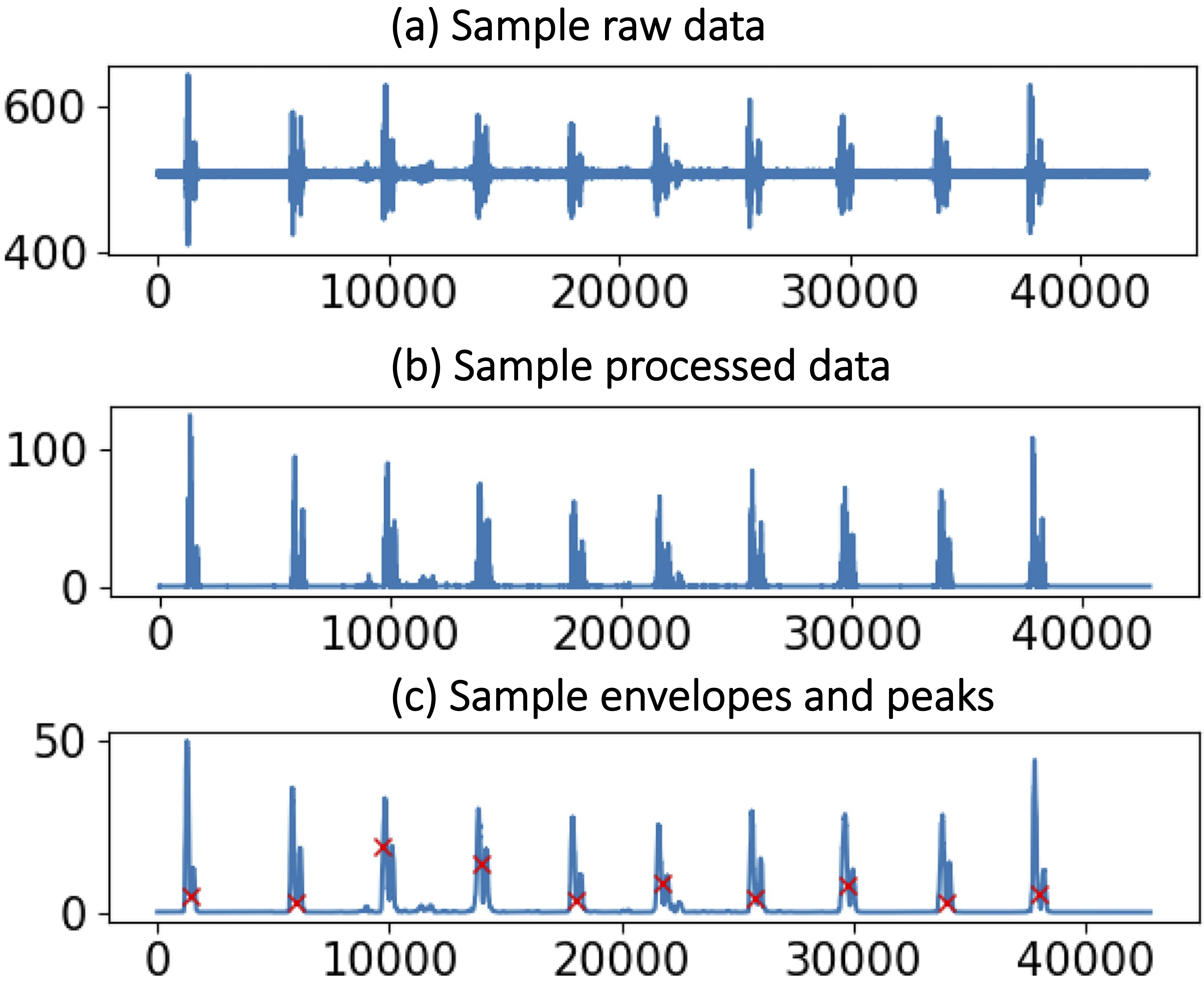}}
\caption{sEMG signal from levator
anguli oris (LAO). (a) Raw data. (b) Processed data before envelope. (c) Envelopes and peaks extracted from processed data.}
\label{Fig:signal_processing}
\end{figure}

The data before and after standard EMG signal processing are shown in Figure~\ref{Fig:signal_processing}. 

\subsection{Word Extraction}
Three steps were involved in word extraction: peak detection, peak localisation, and word localisation.

\textbf{Peak detection:} we detected $10$ peaks corresponding to the $10$ words, ensuring no internal peaks within a single word are included. On the one hand, the interval between peaks was forced to be larger than $3000$ samples ($3$s). On the other hand, if more than $10$ peaks were detected, only the top $10$ most prominent peaks were stored.

\textbf{Peak Location:} notice the peak detection was performed channel-wise. Thus, the three peaks detected from the three channels of a specific word might not be aligned. Hence, peak localisation was conducted to find the mutually recognised peak among the channels. The median value of the three-channel peaks was selected empirically as the mutually recognised peak for a certain word.

\textbf{Word Location:} the peak may not always be the centre of a word signal. Therefore, a sliding window of size $[1500,3]$ ($1500$ samples, $3$ channels) was implemented to localise the word from each of the mutually recognised peaks: the window centre shifted from the $peak-150$ to the $peak+150$ with a $10$ step size, and the centre with the maximum within-window power (sum of the channel power) was selected as the optimal final word centre. The data from the corresponding window was extracted as the final data. 

\begin{figure}[ht!]
\centerline{\includegraphics[width=0.7\columnwidth]{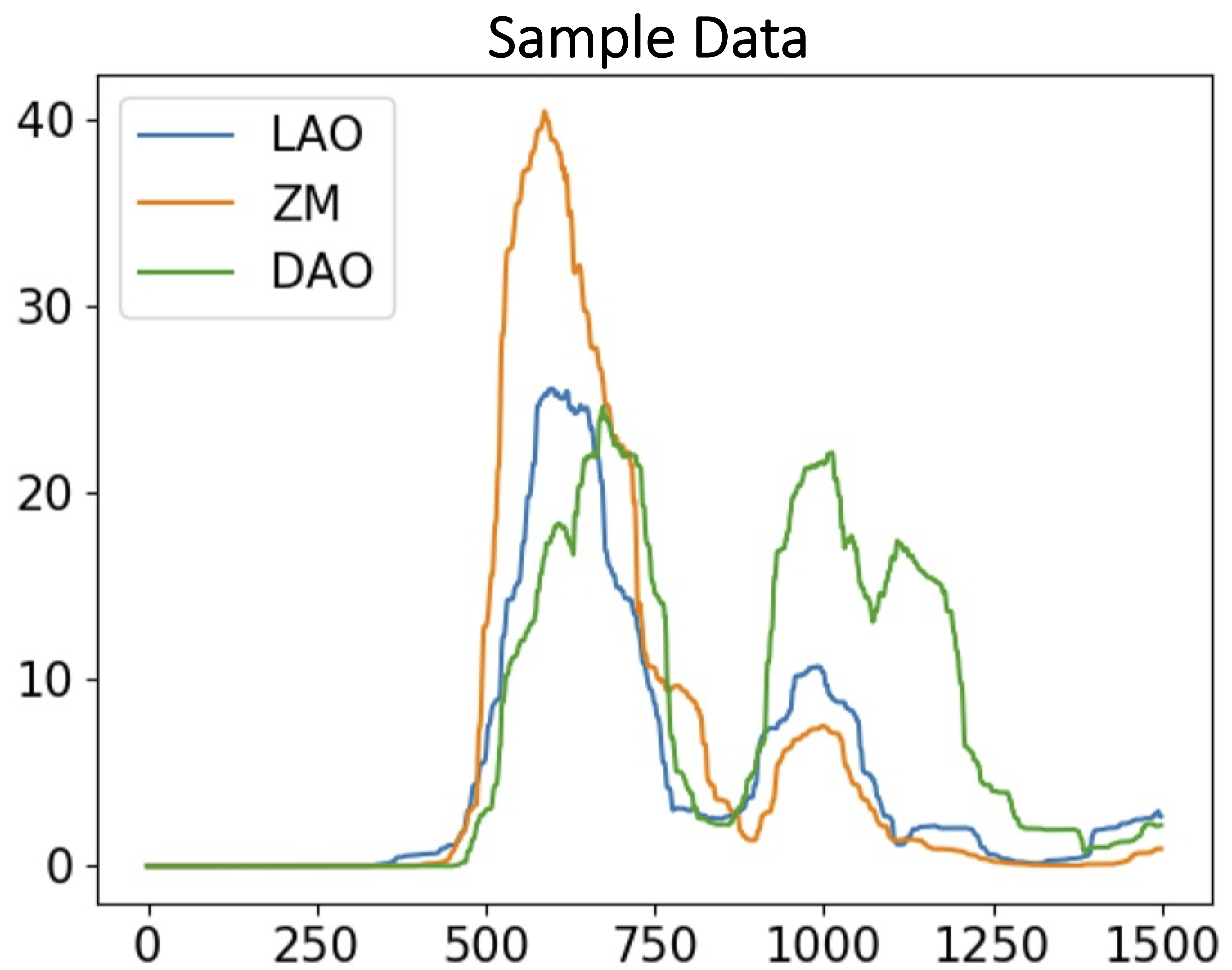}}
\caption{A sample data after processing and extraction pipeline.}
\label{Fig:sample_data}
\end{figure}

The optimal final word centres are shown by the crosses in Figure~\ref{Fig:signal_processing} (c). A sample data is shown in Figure~\ref{Fig:sample_data}.

\section{Method}
\label{sec:Method}
\subsection{Overview}

\begin{figure}[ht!]
\centerline{\includegraphics[width=\columnwidth]{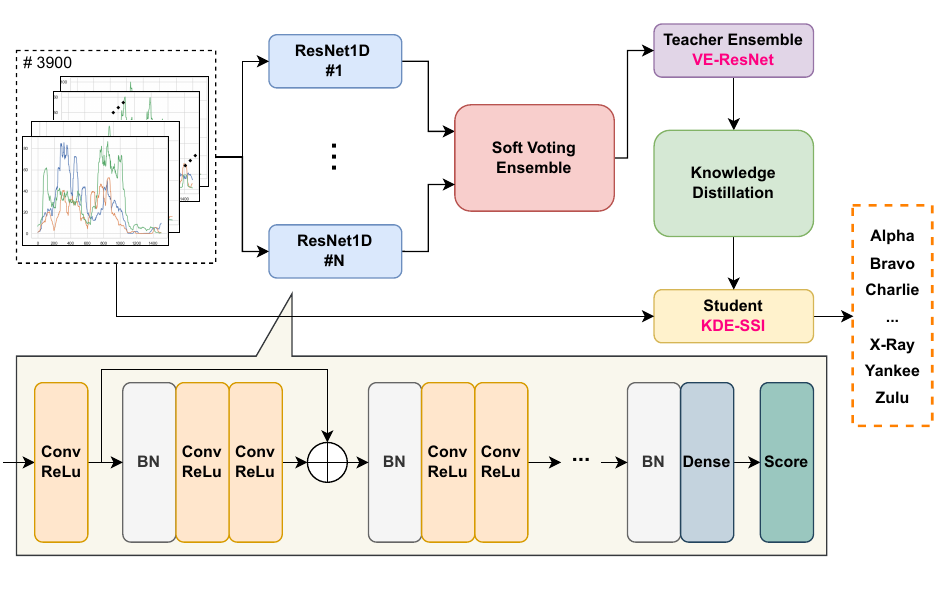}}
\caption{The architecture of KDE-SSI. ResNet1D backbone is shown at the bottom: there are $29$ convolutional layers in total; the Ensemble module is shown in the red block: VE-ResNet takes the probability from the $N$ backbone model and performs a weighted average to get the final ensemble probability; The knowledge distillation module is shown the green boxes: the student model, training with ground truth label using cross-entropy Loss, also mimic the ensemble probability from the teacher VE-ResNet with KL Divergence. }
\label{Fig:KDE}
\end{figure}

The proposed KDE-SSI method is summarised in Figure~\ref{Fig:KDE}. The model contains three parts: the Backbone, the Ensemble module, and the Knowledge Distillation module. The following parts explain each module in detail.

\subsection{Backbone: ResNet1D}
We tested several well-known deep learning architectures as the backbone models. According to the result in Table~\ref{Tab:backbone_perform}, ResNet1D had the best performance and was selected for the ensemble module. 

We implemented a $29$-layer $1D$ ResNet (ResNet1D), which contained $14$ residual blocks, each comprising two $1D$ convolutional layers for extracting time-series features and a \textit{skip connection} to alleviate the degradation problem for the deep network~\cite{resnet}. 
    
\subsection{Ensemble: VE-ResNet}
The voting ensemble is a conventional ensemble learning method for improving model generalisation ability by combining the predictions from multiple models \cite{sagi2018ensemble}. In this study, a voting ensemble ResNet model (VE-ResNet) for the Silent Speech Interface was proposed, which used the ResNet1D as the backbone for training the voting ensemble classifier. In terms of the optimisation procedure, each ResNet1D was trained independently as a base estimator on the same training and validation set. The soft voting strategy (Equation~\ref{Eq:soft_voting}) was applied to merge the predictions from $N$ base estimators $\Theta =\{\Theta_1, \Theta_2, \ldots, \Theta_N \}$ into a final label prediction result. Given the input sEMG signals $\mathbf{x}$, the voting ensemble classifier prediction could be mathematically represented as the following: 
\begin{align}
\label{Eq:soft_voting}
    \mathbf{p_{ve}} &= \sum_{i=1}^{N} w_{i} P\left(\Theta_{i}(\mathbf{x})=c\right) \notag \\ 
    \hat{y} &=\arg \max _{c} \mathbf{p_{ve}}
\end{align}

Where $P\left(\Theta_{i}(\mathbf{x})=c\right)$ indicates the probability that $\Theta_i$ predicts $\mathbf{x}$ belongs to the category $c$, $0\leq w_{i} \leq 1$ denotes the weight of associated with the base estimator $\Theta_{i}$, and $\mathbf{p_{ve}}$ is the final weighted averaged probability output of the VE-ResNet model. $\hat{y}$ is the model prediction, which is the class with the highest probability in $\mathbf{p_{ve}}$.

\subsection{Knowledge Distillation: KDE-SSI}
The classic Knowledge distillation (KD) technique proposed by~\cite{hinton2015distilling}, also known as Vanilla KD, was applied to reduce the complexity of our network as it has been proven effective in reducing the memory footprint and accelerating network inference with minor fluctuations in accuracy~\cite{gou2021knowledge}. Vanilla KD is an offline method requiring a teacher model pre-trained to its optimum and a temperature SoftMax (T-SoftMax) function, which is defined as Equation~\ref{Eq:softmax}, where $z_i$ is the class $i$ input to the T-SoftMax, $T$ is the temperature coefficient, and $p_i$ is the output probability for class $i$. Higher $T$ produces softer labels (smoother probability distribution).
\begin{equation}
    p_i=\frac{exp(\frac{z_i}{T})}{\sum_j{exp(\frac{z_j}{T})}}
    \label{Eq:softmax}
\end{equation}

The loss used to train the student model was the weighted average of two losses (Equation~\ref{Eq:KD_loss}). Minimising the KL divergence loss $D_{KL}$ forced the student model to match the prediction of the teacher model. First, the student logits $\mathbf{z_s}$ and teacher output probability $\mathbf{p_{ve}}$ went through the T-SoftMax with temperature $T$. Then, the $D_{KL}$ loss is calculated as the KL divergence of the two new probabilities. 

The cross-entropy loss $\mathcal{H}$ was added to help when the entropy of soft labels (output of teacher model) was low \cite{hinton2015distilling}. The student logits $\mathbf{z_s}$ went through a standard SoftMax ($T = 1$), and the $\mathcal{H}$ loss was calculated using the SoftMax result and the ground truths $y$ (hard-labels).
\begin{align}
    \mathcal{L}(x;\Theta)&= \alpha T^{2}  \mathcal{D}_{KL}(\sigma(\mathbf{z_s},T),\sigma(\mathbf{p_{ve}},T)) \nonumber\\
    & + (1-\alpha) \mathcal{H}(y,\sigma(\mathbf{z_s},T=1)) 
    \label{Eq:KD_loss}
\end{align}

For the following experiments, we set the weight $\alpha$ of the loss function to $0.5$, so each loss contributes equally.

\subsection{Experimental Setup}
\label{subsec:setup}
The dataset was divided at a ratio of $4$:$1$:$1$, generating a training set, a validation set and a test set, respectively. All evaluated methods were trained using Adaptation Momentum Estimation (Adam) optimizer in $100$ epochs. The early stopping strategy was applied to stop training procedures earlier to prevent unexpected overfitting problems. Models were trained and tested on a Tesla P100 PCIe $16$ GB, and all codes were implemented in Pytorch and available here: \href{https://github.com/laiwenq/AML\_Lymsy}{https://github.com/laiwenq/AML\_Lymsy}.

To verify the superiority of our proposed method, $4$ conventional deep learning methods, including Transformer, ResNet1D, CNN-LSTM, and VGG13, were implemented and compared in terms of precision, recall, F1-score and accuracy. The best-performing architecture was chosen as the backbone for VE-ResNet. Four VE-ResNet consisting of different number of ResNet1D ($N$ = $4$, $6$, $8$ and $10$) were obtained and distilled using different temperatures ($T$ = $5$ and $10$).

\section{Result and Discussion}
\label{sec:result_discussion}
\subsection{Results}
The performance of $4$ evaluated conventional models is summarised in Table~\ref{Tab:backbone_perform}. ResNet1D greatly outperformed the counterparts on the test set among all evaluated methods, achieving the highest scores on all four metrics. VGG13 achieved the second-best performance with an accuracy of $76.1\%$. Although Transformer and LSTM variations were regarded as the gold standard tools for time-series analysis, their four metrics scores were consistently worse than pure CNN-based models (e.g., VGG13, ResNet1D), with at least $15\%$ decrement in accuracy. Since ResNet1D was the most promising among evaluated methods, it was used as the backbone architecture of the proposed voting ensemble classifier.
\begin{table}[ht]
\centering
\caption{Performance of $6$ evaluated methods on $26$ NATO Alphabet classifications. $N$ = $6$ for VE-ResNet and KDE-SSI.}
\begin{tabularx}{1.0\linewidth}{c|c|c|c|c}
\toprule

    Methods                              & Precision (\%) & Recall (\%) & F1-score & Accuracy (\%) \\ \hline \hline
CNN-LSTM  & 57.4           & 56.4        & 0.562    & 56.4          \\ \hline
Transformer & 61.5           & 60.1        & 0.595    & 60.6          \\ \hline
VGG13       & 76.8           & 76.2        & 0.757    & 76.1          \\ \hline
\textbf{ResNet1D}   & \textbf{81.8}          & \textbf{81.0}        & \textbf{0.810}    & \textbf{81.2}    \\ \hline
VE-ResNet & 86.3 & 86.0 & 0.857 & 86.0 \\ \hline
KDE-SSI &  87.4 & 85.7 & 0.855 & 85.9 \\ \bottomrule
\end{tabularx}
\label{Tab:backbone_perform}
\end{table}

As shown in Table~\ref{Tab:KDE}, VE-ResNet with $N$ = $4$ scored the highest before distillation, giving an accuracy of $88.0\%$, while VE-ResNet with $N$ = $6$ and $10$ tied for second place with $86.0\%$ accuracy. After the distillation, all models were decreased in accuracy, and the reduction was inversely related to the value of $T$. 

\begin{table}[ht]
\centering
\caption{Accuracy of VE-ResNet with N estimators ($N$ = $4$, $6$, $8$, $10$).}
\begin{tabularx}{1.0\linewidth}{c|c|c|c}
\toprule
\# of Estimators & VE-ResNet & KDE-SSI (T=5) & KDE-SSI (T=10) \\ \hline \hline
\multicolumn{1}{c|}{4}    &   \textbf{88.0}        &    83.2    &     84.5     \\ \hline
\multicolumn{1}{c|}{\textbf{6}} &      86.0     &  \textbf{85.7}     &    \textbf{85.9}      \\ \hline
\multicolumn{1}{c|}{8}       &     85.7      &   84.6    &    85.2    \\ \hline
\multicolumn{1}{c|}{10}    &     86.0       &   84.9    & 85.7 \\ \bottomrule
\end{tabularx}
\label{Tab:KDE}
\end{table}

At $T$ = $5$, KDE-SSI distilled from $6$ estimators only had a $0.3\%$ of accuracy decrement, while both KDE-SSI with $N$ = $8$ and $10$ had a $1.1\%$ of reduction. Surprisingly, the KDE-SSI distilled from the best-performing VE-ResNet was impacted the most by the distillation, resulting in $4.8\%$ of accuracy loss. 

At $T$ = $10$, the accuracy reduction was attenuated. In particular, KDE-SSI with $N$ = $6$ further approached the accuracy of its teacher model with a negligible accuracy reduction ($0.1\%$).

\begin{figure}[ht!]
    \centering
    \includegraphics[width = 0.85\columnwidth]{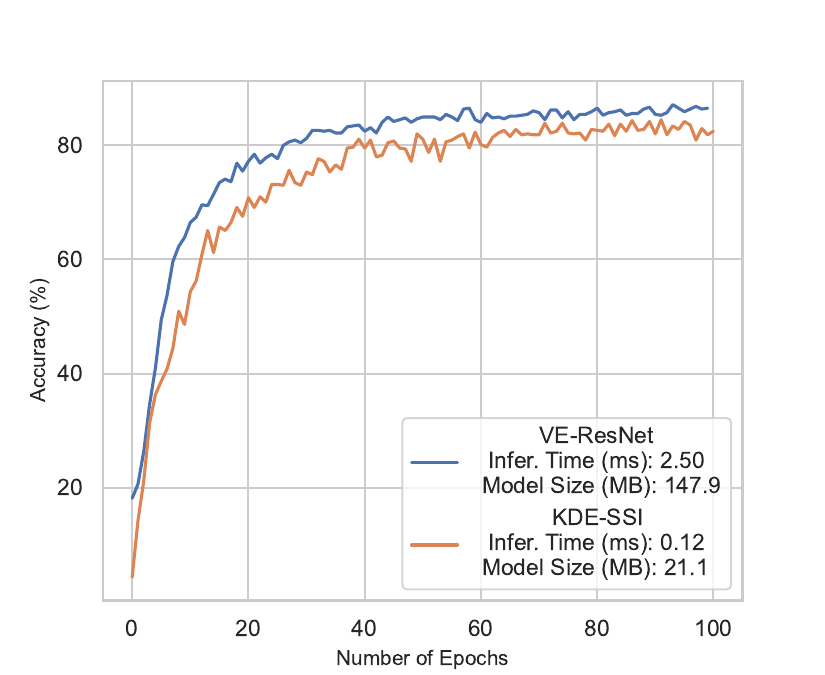}
    \caption{Learning curves of VE-ResNet and KDE-SSI with $N=6$ and $T=10$ on the validation set. The average inference time per sample is $2.50$ms and $0.12$ms, respectively. }
    \label{Fig: learning curve}
\end{figure}
\label{sec:Result}

From the learning curve illustrated in Figure~\ref{Fig: learning curve}, we observed that both VE-ResNet with $N=6$ and its corresponding KDE-SSI distilled with $T=10$ increased their validation accuracy progressively with the number of epochs and did not demonstrate any overfitting problem. 

As expected, a minor accuracy gap between the two models is visible at the end of the training procedure. KDE-SSI was around $20.8$x faster and $7.0$x smaller compared to VE-ResNet. The results in Table~\ref{Tab:best KDE} show that KDE-SSI could generate predictions with at least $0.80$ F1-score for most classes, except the words \textit{Charlie}, \textit{Delta}, \textit{Golf}, \textit{India}, \textit{Quebec}, \textit{Romeo} and \textit{Zulu}, which indicates a significant difference between precision and recall for these words.

\begin{table}[ht]
\caption{Performance of best performing KDE-SSI on 26 NATO Alphabet classification.}
\centering
\begin{tabularx}{0.75\linewidth}{c|c|c|c}
\toprule
\centering
    Class                              & Precision (\%) & Recall (\%) & F1-score  \\ \hline \hline
 Alfa   & 96.4 &  100.0  & 0.982           \\ \hline
 Bravo  &  92.3  & 96.0 &  0.941           \\ \hline
Charlie  &  91.7  & \textbf{52.4} &  \textbf{0.667}          \\ \hline
Delta  &  \textbf{65.7} &  100.0  & \textbf{0.793}          \\ \hline
Echo   & 82.6  &  79.2 &  0.809          \\ \hline
Foxtrot &  96.6 &  96.6 &  0.966          \\ \hline
Golf  &    \textbf{66.7} &  88.9  & \textbf{0.762}          \\ \hline
 Hotel &   95.2 &  83.3  & 0.889          \\ \hline
India   & 93.8  & \textbf{65.2}  & \textbf{0.769}          \\ \hline
Juliett  &  85.7 &  85.7 &  0.857          \\ \hline
Kilo &    90.3  & 96.6  & 0.933           \\ \hline
Lima  &  94.4  & 94.4  & 0.944          \\ \hline
Mike  & 89.7  & 89.7  & 0.897           \\ \hline
November & 100.0 &  87.5  & 0.933          \\ \hline
Oscar  &  87.1 &  90.0  & 0.885          \\ \hline
Papa  & 94.7  & 85.7  & 0.900          \\ \hline
Quebec  &  100.0  & \textbf{60.0}  & \textbf{0.750}          \\ \hline
Romeo  &  \textbf{70.0}  & 84.0  &  \textbf{0.764}          \\ \hline
Juliett  & 85.7 &  85.7  & 0.857          \\ \hline
Sierra  &  100.0  & 76.2  & 0.865          \\ \hline
Tango  &  80.0 &  82.8  & 0.814          \\ \hline
Uniform  &  91.3 &  77.8 &  0.840          \\ \hline
Victor  &  92.3  & 85.7  & 0.889          \\ \hline
Whiskey  &  72.7 &  88.9 &  0.800          \\ \hline
X-ray  &  100.0  & 90.9  & 0.952          \\ \hline
Yankee &   81.8 &  96.4 &  0.885          \\ \hline
Zulu  & \textbf{61.7}  & 95.5 &  \textbf{0.750}          \\
\bottomrule
\end{tabularx}
\label{Tab:best KDE}
\end{table}

\subsection{Discussion}
During the teacher model pre-training stage, we identified the most suitable architecture to fulfil the requirement. As summarised in \cite{Ismail2019}, LSTM-based approaches, such as CNN-LSTM, were unsuitable for long time-series classification problems since they could not efficiently capture long-term temporal dependencies in long sequences. 

The vanilla Transformer performed poorly in our task because the long-range modelling ability of the Transformer was limited by the training data scarcity. Furthermore, the Transformer does not analyse its input sequentially. Using only a conventional absolute position encoding module to learn temporal information in a three-channel sEMG signal might introduce excessive randomness into the attention and limit the model's expressiveness~\cite{ke2020rethinking}.

Pure CNN-based methods (VGG13 and ResNet1D) exhibited far more outstanding performance than their counterparts, which aligned with the findings from \cite{ismail2019deep}. ResNet1D, benefiting from skip connections, outperformed VGG13 in terms of resisting overfitting and gradient vanishing problems. All evaluated models showed similar performance on all metrics, implying that the models tend to give a similar amount of positive and negative predictions. 

Ensembling multiple ResNet1D by soft voting showed a great performance boost in our experiment. The underlying effect of varying the $N$ was not fully understood, as VE-ResNet with the fewest estimators performed the best; however, it had a steep accuracy drop after distillation. A potential explanation was that the soft labels produced by the best VE-ResNet were very distinctive between classes, making the student hard to match. Increasing the $T$ might be a solution since we found that training KDE-SSI with a higher temperature helps the student learn better from the teacher model; however, the convergence would be slower. 

For the best KDE-SSI, we traded negligible accuracy for enormous complexity reduction, making the final model suitable for deployment to the target users. From The inter-class performance of best performing KDE-SSI (Table~\ref{Tab:best KDE}), we observed that well-learned classes exhibited low variance between the performance metrics. For the hard classes, the model tended to prioritise one type of prediction over the other (more positive than negative, or vice versa). This is a sign of poor learning, which could result from multiple factors, such as the inherent lack of enough information. For some words, $3$ facial muscles could not provide sufficient information. For example, people generally do not rely on the chosen muscles to pronounce the word \textit{Charlie}; having more sensors monitoring the tongue or other muscles might be helpful. 

\section{Conclusion}
In this work, we proposed a novel lightweight knowledge-distilled ensemble model for a Silent Speech Interface (KDE-SSI) with a COTS device. We achieved $85.9\%$ accuracy in distinguishing the $26$ NATO phonetic alphabet using sEMG signals collected from $3$ facial muscles. However, KDE-SSI still necessitates a two-stage training procedure (i.e., pre-training and ensembling followed by distillation). Therefore, future work could investigate online distillation approaches capable of simultaneously training a teacher ensemble and student model, resulting in an end-to-end solution for an sEMG-based Silent Speech Interface. Furthermore, domain adaptation approaches could be used to bridge the inter-subject/inter-gender domain shift. A fuzzy matching algorithm could be designed to maintain the usability of KDE-SSI so that KDE-SSI does not necessarily need to recognise every single NATO word correctly. 

\section*{Acknowledgment}
We would like to thank Dr Adam Spiers for providing the BITalino MuscleBIT sEMG sensors used in our research.

\bibliographystyle{IEEEtran}
\bibliography{semg}

\end{document}